# Electric field-induced spin-valley locking in twisted bilayer buckled honeycomb materials

Harold J.W. Zandvliet[1], Pantelis Bampoulis[1], Cristiane Morais Smith[2], and Lumen Eek[2]

[1] Physics of Interfaces and Nanomaterials, MESA+ Institute, University of Twente, P.O. Box 217, 7500AE Enschede, The Netherlands

[2] Institute for Theoretical Physics, Utrecht University, Princetonplein 5, 3584CC Utrecht, The Netherlands

A twisted honeycomb bilayer exhibits a moiré superstructure that is composed of a hexagonal arrangement of AB and BA stacked domains separated by domain boundaries. In the case of twisted bilayer graphene, the application of an electric field normal to the bilayer leads to the opening of inverted band gaps in the AB and BA stacked domains. The inverted band gaps result in the formation of a two-dimensional triangular network of counterpropagating valley protected helical domain boundary states, also referred to as the quantum valley Hall effect. Owing to spin-orbit coupling and buckling, the quantum valley Hall effect in twisted bilayer silicene and germanene is more complex than in twisted bilayer graphene. We found that there is a range of electric fields for which the spin degree of freedom is locked to the valley degree of freedom of the electrons in the quantum valley Hall states, resulting in a stronger topological protection. For electric fields smaller than the aforementioned range the twisted bilayer does not exhibit the quantum valley Hall effect, whereas for larger electric fields the spin-valley locking is lifted and the emergent quantum valley Hall states are only valley-protected.

In the past few decades, topological states of matter have become an intensively studied topic in condensed-matter physics. Of particular interest are topological insulators, which are characterized by an insulating bulk and a metallic surface [1]. In three-dimensional topological insulators, the metallic surface states exhibit the property that the spin of the electron is locked to its momentum. Owing to this spin-momentum locking, the metallic surface states are protected against perfect backscattering. Analogously, in the two-dimensional quantum spin Hall effect (QSH), edge states protected by time-reversal symmetry, enable dissipationless electronic transport, an appealing property for device applications [2-5]. Kane and Mele [4,5] predicted that the honeycomb lattice of graphene could host such QSH edge states as a result of spin-orbit coupling (SOC), which acts as an internal effective magnetic field that pushes spin-up and spin-down electrons in opposite directions. However, graphene's SOC is only on the order of a few µeV, restricting the QSH effect to unrealistically low temperatures. Since the SOC strength scales as $Z^4$, where $Z$ is the atomic number, monoelemental honeycomb materials composed of heavier elements, such as silicene, germanene and stanene, exhibit much larger gaps and thus more robust QSH states [6]. Moreover, these heavier materials favor a buckled honeycomb lattice that is unlike graphene's planar structure, Figure 1a, [7,8]. This buckling allows to tailor the band gap by the application of an electric field. The QSH effect has been recently realized in germanene [9] and germanene nanoribbons [10] at temperatures up to room-temperature [11].

Beyond monolayers, stacking two-dimensional crystals introduces additional degrees of freedom. In particular, a relative twist between layers produces long-wavelength moiré superlattices [12–26], which host a wealth emergent phenomena. A striking example is the realization of unconventional superconductivity in twisted bilayer graphene near the so-called magic angle of approximately 1.1º [13,14]. Within the same system, the application of a perpendicular electric field gives rise to the formation of a two-dimensional network of one-dimensional valley protected helical states located at the domain boundaries between AB and BA stacked regions [27-35]. This phenomenon, known as the valley Hall network, arises from the layer polarization of the valley Chern numbers. In contrast to the spin-momentum locked edge states in QSH systems, these valley-protected states are more fragile, as impurities can induce intervalley scattering.

The principles uncovered in twisted bilayer graphene have inspired the exploration of similar phenomena in other two-dimensional materials. Twisting transition metal dichalcogenides, group-IV honeycomb monolayers, and other van der Waals heterostructures enables the engineering of strongly correlated electronic states, moiré excitons, and topologically nontrivial phases. In these systems, the interplay between SOC, layer hybridization, and moiré potentials provide an even broader platform for realizing and controlling novel topological and correlated quantum states.

In this work, we demonstrate that buckled honeycomb lattices, such as germanene and stanene, with sizeable spin-orbit coupling can restore QSH-like robustness to valley Hall boundary states. In Particular, we find that there is a range of electric fields where the valley degree of freedom gets locked to the spin degree of freedom leading to a two-dimensional triangular network of one-dimensional boundary states that is as strongly protected as the edge states in a QSH insulator.

**Quantum valley Hall effect in twisted bilayer graphene**

In this section, we will first elaborate on the quantum valley Hall effect in small-angle twisted bilayer graphene. Consider a twisted bilayer graphene with a small twist angle (<1º). The moiré periodicity of this twisted bilayer is very large (λ > 15 nm) and the moiré unit cell contains more than 14000 atoms. There are four low-energy bands per moiré unit cell that can contain up to eight electrons. The latter implies that for small twist angles the electron density per unit area is extremely low ($\sim 10^{12}\ cm^{-2}$), and therefore the filling fraction of low-energy bands can easily be varied by relatively small gate voltages [13,14]. The application of an external electric field normal to the bilayer graphene leads to a charge shift from an A (B) atom to a B (A) atom and thus to the opening of inverted band gaps in the AB and BA stacked domains [36-37]. Therefore, topologically protected states will emerge at the boundaries between the AB and BA stacked domains, see Figure 1(b) [27-35]. The number of domain boundary states is given by the Chern number. The Chern number can be obtained by integrating the Berry curvature over the first Brillouin zone. When a perpendicular electric field is applied, the inversion symmetry is broken, resulting in opposite Chern valley numbers for the BA and AB stacked domains. Since the valley Chern numbers change from +1 (AB domain) to -1 (BA domain) and from -1 (AB) to +1 (BA) at the *K* and *K′* valleys, respectively, the total number of conducting channels is four (two per valley, since $C_K = C_K^{AB} - C_K^{BA} = 1 - (-1) = 2$ and $C_{K'} = -2$) resulting in a total conductance of $4e^2/h$ [28-31,38]. Unfortunately, the topological protection is rather weak as the presence of defects or impurities give rise to intervalley scattering, which destroys, at least locally, the topological protection.

The idea to exploit these valley-protected channels that propagate along the AB/BA boundaries of bilayer graphene has been pursued by a few groups. By using a split gate to produce opposite transverse electric fields Martin *et al.* [39] were the first to demonstrate that a one-dimensional conducting channel in conventional Bernal stacked bilayer graphene can indeed be realized. The experimental realization of a split gate is, however, a quite challenging task. Alternatively, one could use a single gate and try to 'find' regions with different stacking orders. Lu *et al.* [40] pursued this idea and managed to identify a few of these valley-polarized topologically protected conducting channels. A twisted bilayer is a much more appealing system to study the quantum valley Hall effect because the stacking order changes continuously over the whole bilayer, and therefore this system results in a 2D triangular network of AB/BA domains boundaries.

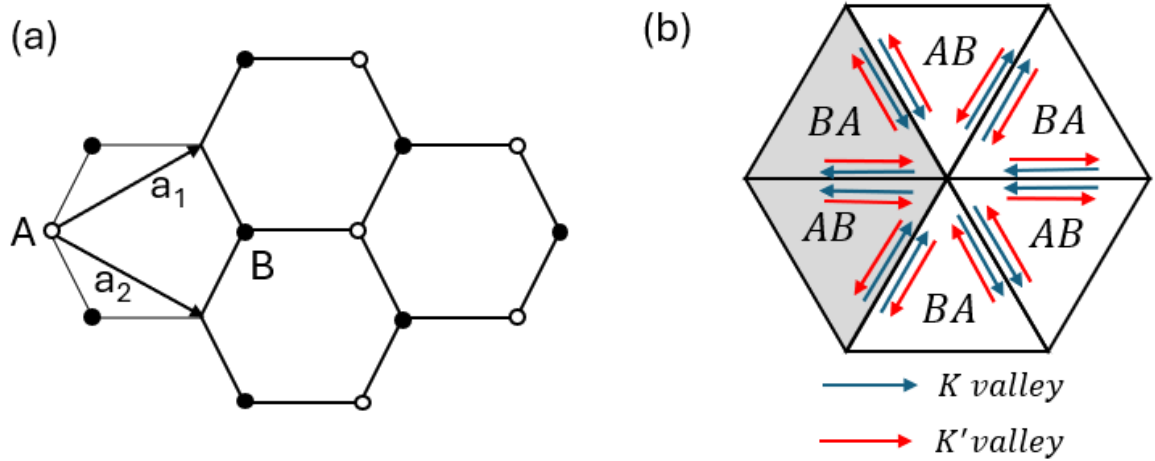

*Figure 1* (a) *Schematic diagram of the buckled honeycomb lattice of silicene and germanene. The unit cell is spanned by lattice vectors $a_1$ and $a_2$ and involves an upward buckled atom (A) and a downward buckled atom (B). (b) 2D network of valley protected domain boundary states in a twisted bilayer. The moiré unit cell (grey area) consists of a triangular AB domain and a triangular BA domain.*

**Quantum valley Hall effect in buckled honeycomb bilayer materials**

As already mentioned before there are some differences between graphene and buckled honeycomb materials, such as silicene and germanene. Silicene and germanene have a larger spin-orbit coupling than graphene, their honeycomb lattices are buckled, the intralayer coupling is weaker and the interlayer coupling is stronger than in graphene. Despite the differences, the electronic band structure of bilayer silicene and germanene is qualitatively very similar to bilayer graphene.

We first consider a domain wall between an infinitely large AB and BA stacked region. In Figures 2(a)-(b) we show a structural model of an AB and BA stacked buckled bilayer. The buckling of the two layers is assumed to be in-phase, that is, in both layers the A sublattice buckles upward, while the B sublattice buckles downward. The Hamiltonian of an AB stacked region reads,

$$H = t \sum_l \sum_{\langle i_l j_l \rangle} c_{i_l}^\dagger c_{j_l} + t_\perp \sum_{l \neq l'} \sum_{\langle i_l j_{l'} \rangle} c_{i_l}^\dagger c_{j_{l'}} + \frac{i\lambda_{SO}}{3\sqrt{3}} \sum_l \sum_{\langle\langle i_l j_l \rangle\rangle} v_{i_l j_l} c_{i_l}^\dagger s^z c_{j_l} + U_\perp \sum_l \sum_{i_l} \chi_l c_{i_l}^\dagger c_{i_l}$$
$$+ M \sum_l \sum_{i_l} \xi_{i_l} c_{i_l}^\dagger c_{i_l}.$$

(1)

Here, $c_{i_l}^\dagger$ ($c_{i_l}$) is the electronic creation (annihilation) operator of an electron on site $i$ of layer $l$. The first two terms describe nearest-neighbor hopping with strength $t$ and $t_\perp$ for intralayer and interlayer hopping parameters, respectively. The third term describes intralayer spin-orbit coupling with strength $\lambda_{SO}$. Here, $v_{ij} = (\hat{d}_1 \times \hat{d}_2)_z$, with $\hat{d}_1$ and $\hat{d}_2$ the unit vectors along the two bonds the electron traverses by going from site $i$ to $j$, and $s^z$ is the Pauli z-matrix acting on spin. $\langle i_l j_l \rangle$ and $\langle\langle i_l j_l \rangle\rangle$ refers to a summation over nearest and next-nearest neighbors, respectively. The last two terms describe onsite terms, with $U_\perp$ the interlayer potential difference and $M$ the sublattice potential difference. Consequently, $\chi_l = 1$ ($-1$) for the top (bottom) layer and $\xi_i = 1$ ($-1$) for the A (B) sublattice. There are in principle two spin-orbit terms: the intrinsic spin-orbit coupling $\lambda_{SO}$ and the Rashba spin-orbit coupling $\lambda_R$. For monolayer systems, the latter term vanishes at the $K$ and $K'$ points, and can therefore safely be ignored [41]. However, for Bernal bilayers, the layer resolved Rashba spin-orbit coupling can have a nonzero value [42]. For the sake of simplicity, we consider here only the intrinsic spin-orbit coupling term, $\lambda_{SO}$.

Upon expanding around the $K$ and $K'$ point (in the absence of spin-orbit coupling, or buckling), there are four spin-degenerate energy bands that are given by [43,44],

$$E_{\alpha=1,2}^\pm = \pm\sqrt{A(k) + (-1)^\alpha \sqrt{B(k)}} \tag{2a}$$

with

$$A(k) = \frac{t^2}{2} + \frac{U_\perp^2}{4} + \left(v_F^2 + \frac{v_\perp^2}{2}\right)\hbar^2 k^2 \tag{2b}$$

$$B(k) = \frac{(t^2 - v_\perp^2 \hbar^2 k^2)^2}{4} + \hbar^2 v_F^2 k^2 (t^2 + U_\perp^2 + \hbar^2 v_\perp^2 k^2) + 2t\xi v_{\perp,1} v_F^2 \hbar^3 k^3 \cos(3\theta) \tag{2c}$$

where $\boldsymbol{k} = (k\cos\theta, k\sin\theta)$ is the wavevector in the vicinity of the $K$ ($\xi = 1$) and $K'$ ($\xi = -1$) points, $k = |\boldsymbol{k}|$ and $v_F = (\sqrt{3}/2\hbar)at$ [$v_\perp = (\sqrt{3}/2\hbar)at_\perp$] is the in-plane [out-of-plane] Fermi velocity.

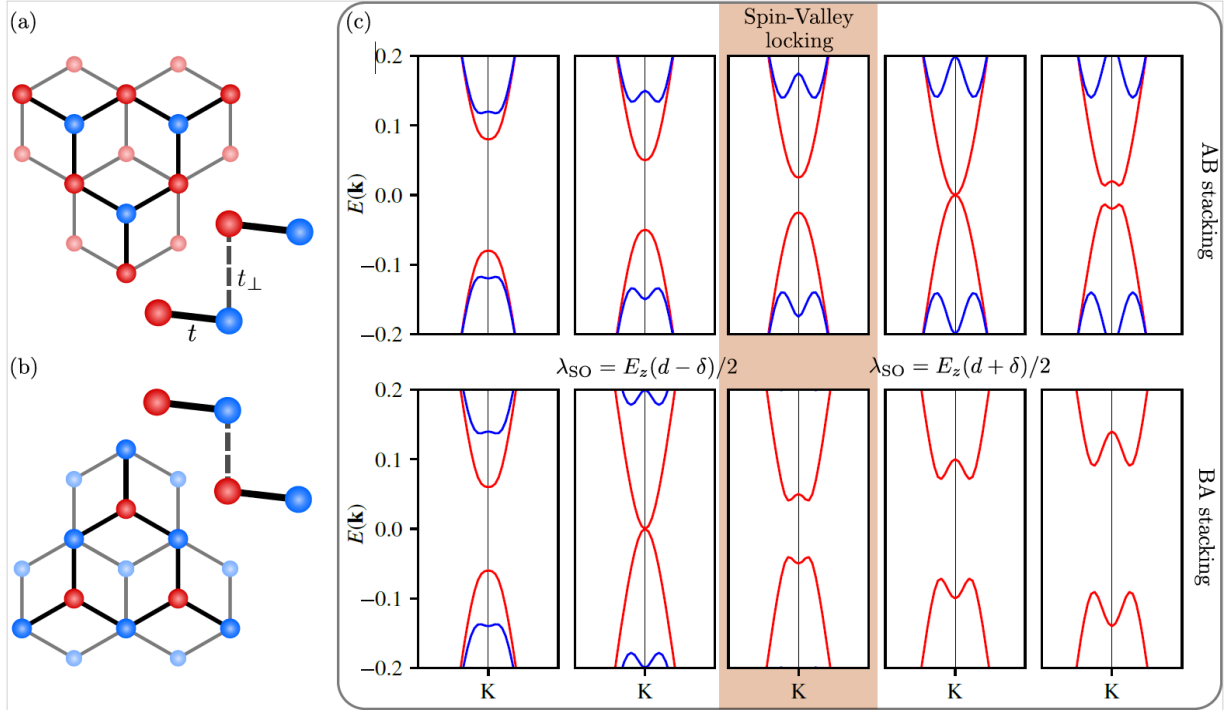

Figure 2 (a) –(b) Structural model of AB and BA stacked bilayer germanene. Upward buckled atoms are red and downward buckled atoms are blue (c) Low-energy electronic band structure of an AB (top) and BA (bottom) stacked honeycomb bilayer; red and blue indicate spin up and down, respectively. All panels have $\lambda_{SO}/t = 0.1$ and $t_\perp/t = 0.3$.

Furthermore, we have (from left to right): $U_\perp/t = (0.03, 0.075, 0.1125, 0.15, 0.18)$ and $M/t = (0.01, 0.025, 0.0375, 0.05, 0.06)$. For $U_\perp - M \leq \lambda_{SO} \leq U_\perp + M$ there is spin-valley locking.

The application of an interlayer bias $U_\perp$ normal to the bilayer results in a spin-splitting of the low-energy bands. In the remainder of this paragraph, we will elaborate on the effect of an electric field on the spin-splitting of the low-energy bands, and we will derive the threshold bias that is required to open an inverted band gap in AB and BA stacked domains. The buckling allows to tailor the size of the band gap using an applied electric field [9,45,46]. In this case one has,

$$U_\perp = \frac{d}{2}E_z, \qquad M = \frac{\delta}{2}E_z,$$

(3)

where $d$ is the interlayer separation, $M$ the intralayer bias, $\delta$ the buckling and $E_z$ the applied out of plane electric field. The application of an electric field, which we assume here to be positive, in a direction normal to the bilayer, results in different band gaps for the spin up and spin down bands. We assume that the electric field is positive and smaller than the critical electric field for gap closure. At the $K$-point ($\xi = 1$), the spin up and spin down band respond in a different way to the applied electric field. For small electric fields, the band gaps in the AB and BA stacked domains are not inverted and therefore there are no topologically protected domain boundary states [47]. An inverted band gap opens up at the $K$ point ($\xi = 1$) in the AB and BA stacked regions if,

$$\xi = 1, s = 1: \qquad (\lambda_{SO} - U_\perp) \mp M = 0, \qquad E_{z,c} = \frac{2\lambda_{SO}}{\delta \pm d} \qquad (4a)$$

$$\xi = 1, s = -1: \qquad (U_\perp + \lambda_{SO}) \pm M = 0, \qquad E_{z,c} = \frac{2\lambda_{SO}}{\mp d - \delta} \qquad (4b)$$

where $\mp$ denotes AB/BA stacking. If we consider positive $E_z$ and $\lambda_{SO}$, the gap closing at the $K$ point occurs for the spin-up band. First, the 'inverted' spin up band gap at the $K$ point opens up at a critical field $E_{z,c1} = 2\lambda_{SO}/(d + \delta)$ in the BA region. Upon increasing the electric field further, this band inversion of the spin-up band also occurs in the AB region for $E_{z,c2} = 2\lambda_{SO}/(d - \delta)$. In Figures 2(c), the low-energy band structure of an AB and BA stacked bilayers system at the $K$ point is plotted for different electric fields. At the $K'$ point, the same happens but for the spin-down band.

In the intermediate region, i.e. $E_{z,c1} \leq E_z \leq E_{z,c2}$, there is band inversion for the spin-up band at $K$ in the BA region, but not in the AB region. When the two regions are interfaced, there will be a topological spin-up state with valley quantum number $K$. Likewise, there is a spin-down state with quantum valley number $K'$.

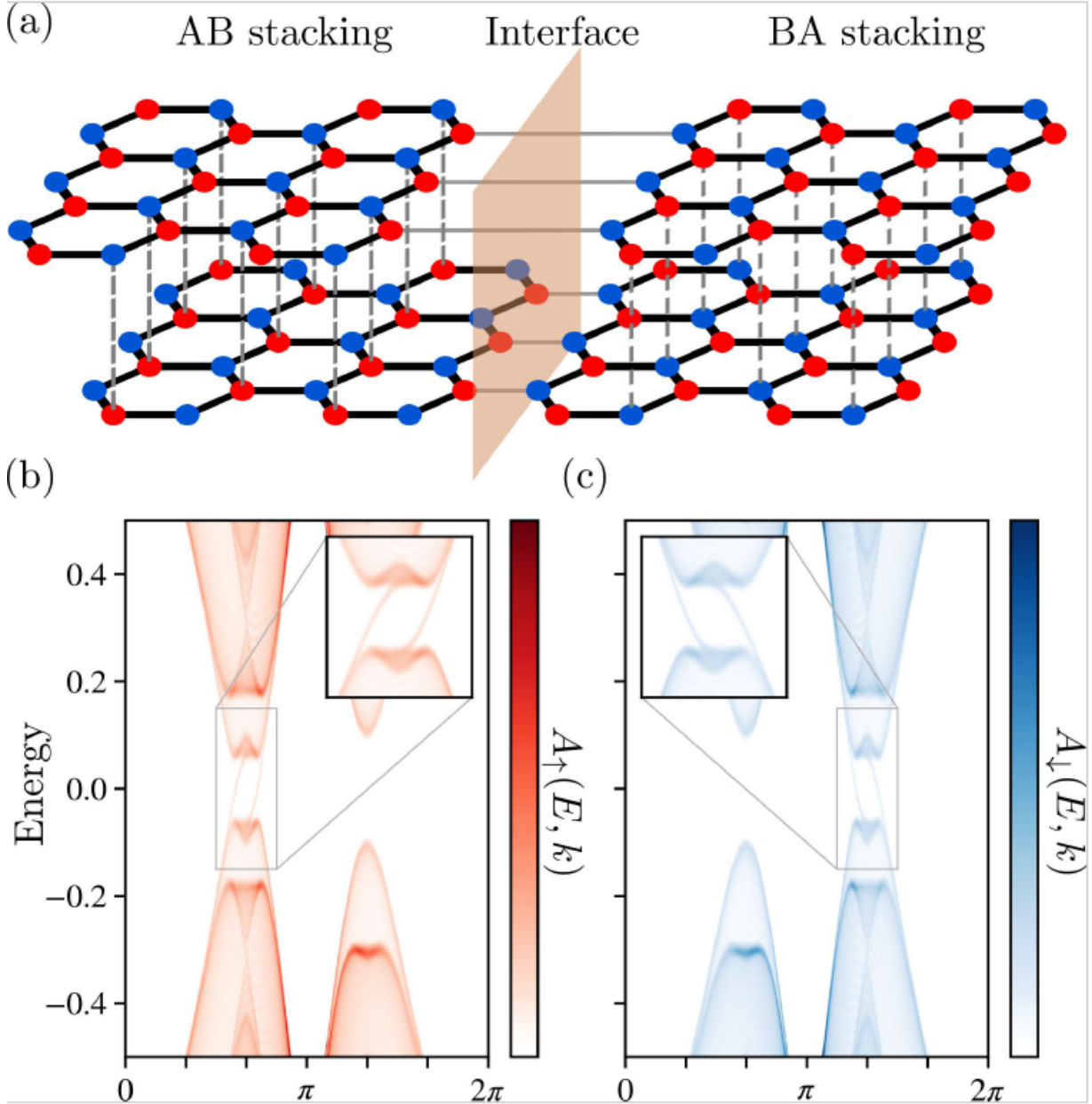

*Figure 3* *(a) Structural model of an AB/BA domain in a Bernal bilayer (b)-(c) Tight-binding calculations of the spin-resolved interface spectral function, $A_\sigma(E,k)$, at the K and K' points for an electric field in the interval $[E_{z,c1}, E_{z,c2}]$. Here, we used $t_\perp = 0.3t$, $\lambda_{SO} = 0.2t$, $U_\perp = 0.2t$, and $M = 0.05t$.*

To better illustrate this behavior, we model an interface between an AB and BA stacked region in a similar manner to the set-up proposed by Vaezi *et al.* [48], see Figure 3(a). Here, we connect the atoms over the interface artificially with the same hopping strength $t$. While the geometry in Figure 3(a) is artificial, it correctly models the behavior of an AB/BA interface which occurs, for example, in strained bilayer or twisted bilayer moiré systems [27-30]. We consider the interface between two large AB and BA regions and calculate the spin-resolved interface spectral function $A_\sigma(E, k)$, see Figures 3(b)-(c). In the critical region, we observe a pair of right propagating spin-up interface states at the $K$ valley, while there is a pair of left propagating spin-down states at the $K'$ valley.

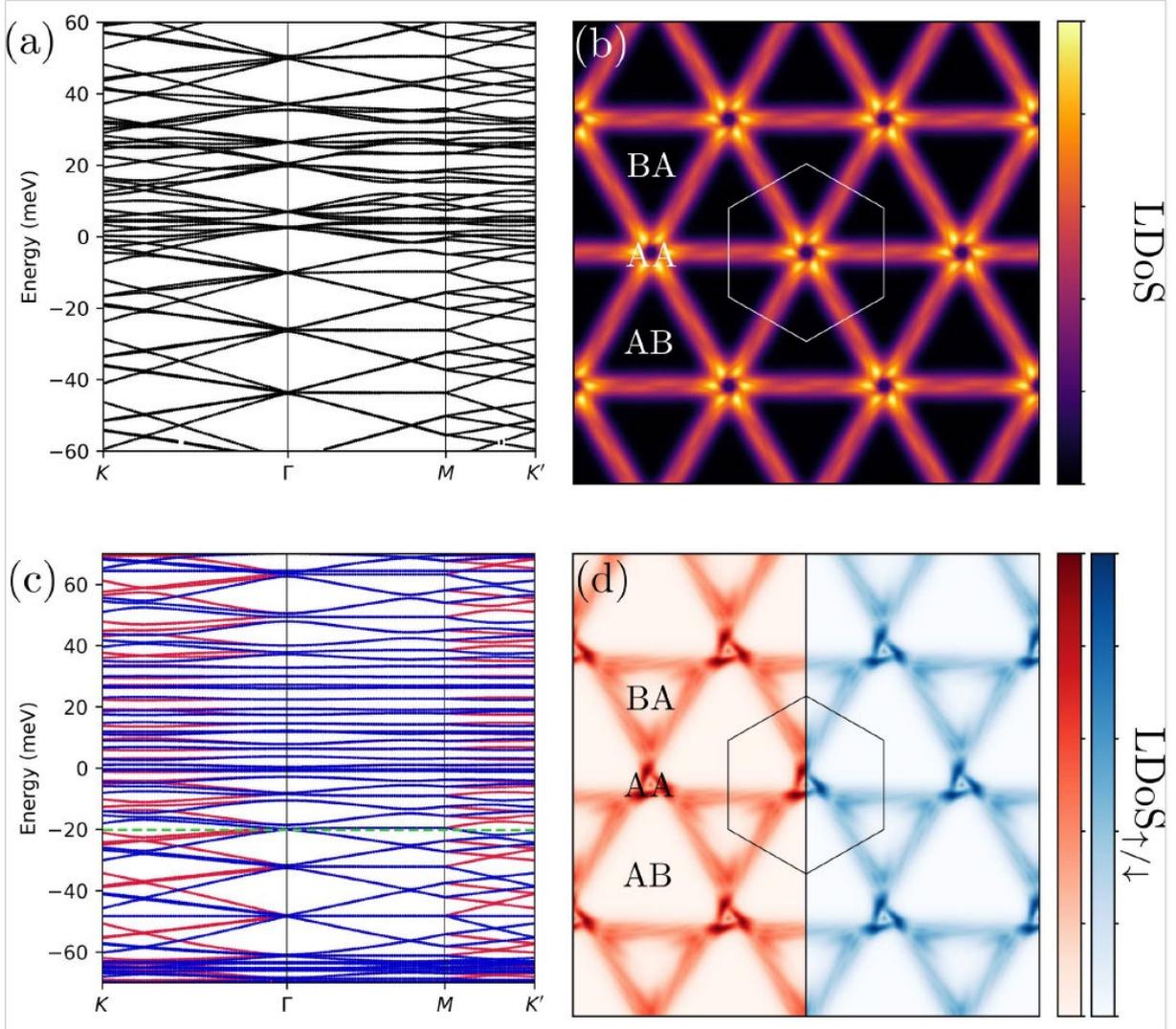

*Figure 4:* *Valley Hall networks in twisted bilayer honeycomb systems. (a) Low energy spectrum of a relaxed unbuckled honeycomb bilayer with a twist angle of $\theta = 0.47°$ for $\lambda_{SO} = M = 0$ and $U_\perp = 200\ meV$. (b) LDoS corresponding to the orange line in (a). High symmetry points and the moiré unit cell have been indicated. (c) Low energy spectrum of a relaxed buckled honeycomb bilayer with a twist angle of $\theta = 0.47°$ for $\lambda_{SO} = 0.25\ eV$, $U_\perp = 250\ meV$, and $M = 100\ meV$. Red corresponds to spin up and blue corresponds to spin down. (d) Spin-resolved LDoS corresponding to the green line in (c), left shows spin up while right shows spin down. High symmetry points and the moiré unit cell have been indicated.*

**Twisted bilayers**

Interfaces between AB and BA stacked regions naturally occur within the moiré unit cells of (small angle) twisted bilayer systems. To model this, we consider the Kane-Mele Hamiltonian on a twisted bilayer system:

$$H = t \sum_l \sum_{\langle i_l j_l \rangle} c^\dagger_{i_l} c_{j_l} + \sum_{l \neq l'} \sum_{\langle i_l j_{l'} \rangle} t_\perp(\mathbf{r}_{i_l} - \mathbf{r}_{j_{l'}}) c^\dagger_{i_l} c_{j_{l'}} + \frac{i\lambda_{SO}}{3\sqrt{3}} \sum_l \sum_{\langle\langle i_l j_l \rangle\rangle} \nu_{i_l j_l} c^\dagger_{i_l} s^z c_{j_l}$$
$$+ U_\perp \sum_l \sum_{i_l} \chi_l c^\dagger_{i_l} c_{i_l} + M \sum_l \sum_{i_l} \xi_{i_l} c^\dagger_{i_l} c_{i_l}.$$

(5)

Here, we parametrize the interlayer hopping through the function,

$$t_\perp(\mathbf{r}_1 - \mathbf{r}_2) = t_\perp \cdot \left[1 - \left(\frac{(\mathbf{r}_1 - \mathbf{r}_2) \cdot \mathbf{e}_z}{|\mathbf{r}_1 - \mathbf{r}_2|}\right)^2\right] \exp\left\{-\frac{|\mathbf{r}_1 - \mathbf{r}_2| - a_{cc}}{\delta}\right\},$$

(6)

where $t_\perp$ is the interlayer hopping strength, $d$ and $a_{cc}$ are the layer separation and the atom-atom distance in the monolayers, $\mathbf{e}_z$ is the unit vector in the z-direction and $\delta$ is a parameter that determines the decay of the hopping function. In the remainder of this manuscript, we take $\delta = 0.05\, a_{cc}$ and $t_\perp = 0.4\, eV$.

The twist angles required to obtain valley Hall networks are usually exceedingly small, as it requires a sufficiently large AB/BA regio to exist within the moiré unit cell. Ref. [30] gives a condition for the maximum twist angle,

$$144(\ln 0.05)^2 \left(\frac{\hbar v_F \theta}{t_\perp a_0}\right)^2 < \frac{U_\perp}{t_\perp},$$

(7)

which yields approximately θ ≈ 0.2 − 0.3° for graphene parameters [49,50]. The gap that allows for spin-valley locking is generally smaller than the valley-Hall gap and, therefore, a spin-valley network requires a smaller twist angle or stronger interlayer hopping (recall that ℓ scales inversely with gap size). However, this condition is made under the assumption that the bilayer system is unrelaxed. For small twist angles in graphene, relaxation takes place such that the AB and BA regions grow larger in area, at the cost of the size of the AA region [51]. As a consequence of the larger AB/BA regions in relaxed unit cells, (spin-locked) valley Hall network will show up already at larger twist angles. This makes it numerically tractable to obtain the dispersion relation and (spin-resolved) local density of states (LDoS) for such systems. In app. A, we explain further how the atomic positions of a relaxed bilayer are obtained. On this relaxed lattice we then apply the bilayer Kane-Mele Hamiltonian cf. Eq. (5).

Figure 4a, depicts the low-energy dispersion relation for an unbuckled honeycomb bilayer with a twist angle of $\theta = 0.47°$. Here we use the parameters $t = -1\ eV$, $t_\perp = 0.4\ eV$ and $\delta = 0.05$. Furthermore, we have $\lambda_{SO} = M = 0$ and $U_\perp = 200\ meV$. Due to the relaxation, the characteristic flat bands around $E = 0\ meV$ have partially disappeared. Nevertheless, at higher/lower energies, dispersive states are clearly visible. Figure 4b, shows the LDoS at $E = -30\ meV$ (indicated by the orange line in Figure 4a). We have indicated the different atomic registries and the moiré unit cell with white lines. On the interface of the AB and BA regions the valley Hall network can be observed.

Next, we investigate the spin-valley locked phase. To this extent, we take $\lambda_{SO} = 0.25\ eV$, $U_\perp = 0.25\ eV$, and $M = 0.1\ eV$. Figure 4c shows the low-energy dispersion relation for these parameters. The spin-degeneracy that was present in Fig 4a has now been lifted due to the combination of spin-orbit coupling and the interlayer bias. Finally, we show an example LDoS (taken at $E = -20\ meV$, depicted by the green line in Fig. 4c) of a spin-valley network in Fig 4(d) for a buckled bilayer with twist angle $\theta = 0.47°$, the same hopping parameters as in (a) and $U_\perp = 250\ meV$, $M = 100\ meV$. Here, the left half of the panel shows the spin-up resolved LDoS while the right half shows the spin-down resolved LDoS. While these LDoS look identical for spin up and down, respectively, they represent counterpropagating states, which can be deduced from the spectrum in Fig. 4c. Notice that, while for the `normal' quantum valley Hall network, the states sit exactly at the interface of AB and BA regions, for the spin-valley locked network, there is an asymmetry. This is caused by the broken symmetry in the bilayer due to the buckling, where AB and BA regions are structurally now different, (see e.g. Figures 2a and 2b).

**Relevance of spin-valley locking**

The spin-valley locking has two important implications for the topological protection and the conductance of the domain boundaries states. Firstly, the quantum valley Hall boundary states are as robust as the quantum spin Hall edge states. Secondly, the conductance of the domain boundary states is not $4e^2/h$, but only $2e^2/h$. For electric fields exceeding $\frac{2\lambda_{SO}}{(d-\delta)}$ the spin-valley locking is lifted; the boundary states are only valley protected and the conductance increases to $4e^2/h$ The valley protection can be lifted by intervalley scattering, which can be induced by scattering at defects or impurities [34].

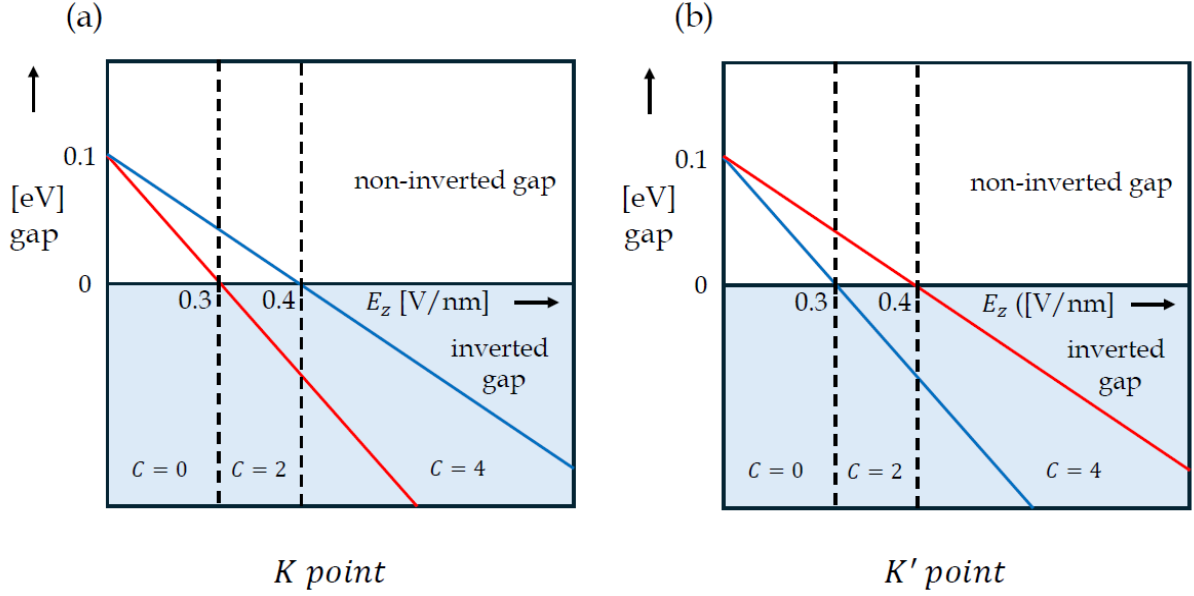

*Figure 5* Phase diagram for small-angle twisted bilayer germanene at the K point (a) and K' point (b) as a function of an applied electric field. Red and blue lines refer to spin up and spin down bands, respectively. C is the Chern number. In the shaded blue area the gap is inverted. ($\lambda_{SO} = 50\ meV$, $\delta = 0.04\ nm$ and $d = 0.28\ nm$).

Our results are applicable to 2D materials like silicene and germanene. These materials have a sizeable spin-orbit coupling and are buckled. In Figure 5 we show a phase diagram for small angle twisted bilayer germanene. When $\lambda_{SO} = 50\ meV$, $\delta = 0.04\ nm$ and $d = 0.28\ nm$, we find a critical electric field for spin-valley locking that ranges from about 0.3 V/nm to about 0.4 V/nm. These critical fields can easily be achieved by conventional gating.

To date several groups have already managed to realize twisted bilayer silicene and germanene [52-54]. The band gap in the AB and BA stacked domains of twisted bilayer germanene can be varied by an electric field, but unfortunately the twist angle was too large for the observation of the quantum valley Hall effect. Since silicene and germanene are not stable under ambient conditions, the twisted bilayers must be grown or synthesized at ultra-high vacuum conditions. So far, two methods have been explored. In the first method, multilayer silicene or germanene is grown on a suitable substrate. Although the AA stacking is preferred for bilayer silicene and germanene, one also finds twisted regions [51-54]. In the second method, part of the top layer of a monolayer or multilayer of silicene or germanene is flipped using the tip of a scanning tunneling or atomic force microscope. This method, which has been introduced by Chen *et al.* [55], has already been successfully applied to graphene.

**Conclusions**

We have shown that the spin-orbit coupling and buckling in twisted bilayer silicene and germanene results in a quantum valley Hall effect that is more complex than the quantum valley Hall effect in twisted bilayer graphene. For small electric fields, the twisted bilayer does not possess inverted band gaps and there are no quantum valley Hall states. For electric fields in the range of $[E_{z,c1}, E_{z,c2}]$, inverted band gaps in the AB and BA stacked domains emerge, but the valley degree of freedom of the electrons in domain boundary states is locked to the spin degree of freedom. This results in a stronger topological protection and a reduction of the number of domain boundary channels by a factor of 2. For electric fields exceeding $E_{z,c2}$, this spin-valley locking is lifted and a triangular 2D network of four valley protected helical domain boundary states develops.

**Acknowledgements**

H.J.W.Z, C.M.S., and L.E acknowledge the research program "Materials for the Quantum Age" (QuMat) for financial support. This program (registration number 024.005.006) is part of the Gravitation program financed by the Dutch Ministry of Education, Culture and Science (OCW). P.B. acknowledges funding from the European Union and the Dutch Research Council (NWO, grant numbers: OCENW.M22.123 and NWO Vidi VI.Vidi.233.019). Funded by the European Union (ERC, Q-EDGE, 101162852). Views and opinions expressed are however those of the author(s) only and do not necessarily reflect those of the European Union or the European Research Council. Neither the European Union nor the granting authority can be held responsible for them.

# References

[1]. M.Z. Hasan and C.L. Kane, *Colloquium: Topological Insulators,* Rev. Mod. Phys. **82**, 3045 (2010).

[2]. B.A. Bernevig, T.L. Hughes, and S.-C. Chang, *Quantum spin Hall effect and topological phase transition in HgTe quantum wells*, Science **314**, 1757 (2006).

[3]. B.A. Bernevig, and S.-C. Chang, *Quantum spin Hall effect*, Phys. Rev. Lett. **96**, 10682 (2005).

[4]. C. L. Kane and E. J. Mele, *Quantum Spin Hall effect in Graphene*, Phys. Rev. Lett. **95**, 226801 (2005).

[5]. C. L. Kane and E. J. Mele, *$Z_2$ Topological Order and the Quantum Spin Hall Effect*, Phys. Rev. Lett. **95**, 146802 (2005).

[6]. C.-C. Liu, W. Feng, and Y. Yao, *Quantum Spin Hall Effect in Silicene and Two-Dimensional Germanium*, Phys. Rev. Lett. **107**, 076802 (2011).

[7]. M. Houssa, A. Dimoulas and A. Molle, *Silicene: a review of recent experimental and theoretical investigations*, J. Phys. Cond. Mat. **27**, 253002 (2015).

[8]. A. Acun, L. Zhang, P. Bampoulis, M. Farmanbar, M. Lingenfelder, A. van Houselt, A.N. Rudenko, G. Brocks, B. Poelsema, M.I. Katsnelson and H.J.W. Zandvliet, *Germanene: the germanium analogue of graphene*, J. Phys. Cond. Mat. **27**, 443002 (2015).

[9]. P. Bampoulis, C. Castenmiller, D.J. Klaassen, J. van Mil, Y. Liu, C.-C. Liu, Y. Yao, M. Ezawa, A.N. Rudenko and H.J.W. Zandvliet, *Quantum Spin Hall States and Topological Phase Transition in Germanene*, Phys. Rev. Lett. **130**, 196401 (2023).

[10]. D.J. Klaassen, L. Eek, A.N. Rudenko, E.D. van 't Westende, C. Castenmiller, Z. Zhang, P.L. de Boeij, A. van Houselt, M. Ezawa, H.J.W. Zandvliet, C.M. Smith, and P. Bampoulis, Realization of a one-dimensional topological insulator in ultrathin germanene nanoribbons, *Nature Communications* **16**, 2059 (2025).

[11]. D.J. Klaassen, I. Boutis, C. Castenmiller and P. Bampoulis, *Tunability of topological edge states in germanene at room temperature*, J. Mat. Chem. C **12**, 15975 (2024).

[12]. R. Bistritzer and A. MacDonald, *Moiré butterflies in twisted bilayer graphene,* Phys. Rev. B **84**, 035440 (2011).

[13]. Y. Cao, V. Fatemi, A. Demir, S. Fang, S.L. Tomarken, J.Y. Luo, J.D. Sanchez-Yamagishi, K. Watanabe, T. Taniguchi, R.C. Ashoori and P. Jarillo-Herrero, *Correlated insulator behaviour at half-filling in magic-angle graphene superlattices*, Nature **556**, 80 (2018).

[14]. Y. Cao, V. Fatemi, S. Fang, K. Watanabe, T. Taniguchi, E. Kaxiras and P. Jarillo-Herrero, *Unconventional superconductivity in magic-angle graphene superlattices*, Nature **556**, 43 (2018).

[15]. X. Lu et al., *Superconductors, orbital magnets and correlated states in magic-angle bilayer graphene*, Nature **574**, 653 (2019).

[16]. P. Stepanov et al., *Untying the insulating and superconducting orders in magic-angle graphene*, Nature **583**, 375 (2020).

[17]. G. Li, A. Luican, J. M. B. Lopes dos Santos, A. H. Castro Neto, A. Reina, J. Kong and E.Y. Andrei, *Untying the insulating and superconducting orders in magic-angle graphene,* Nature Phys. **6**, 109 (2010).

[18]. L. Van Hove, *The Occurrence of Singularities in the Elastic Frequency Distribution of a Crystal,* Phys. Rev. **89**, 1189 (1953).

[19]. T. Ohta, A. Bostwick, T. Seyller, K. Horn and E. Rotenberg, *Controlling the Electronic Structure of Bilayer Graphene,* Science **313**, 951 (2006).

[20]. E.V. Castra, K.S. Novoselov, S.V. Morozov, N.M.R. Peres, J.M.B. Lopes dos Santos, J. Nilsson, F. Guinea, A.K. Geim and A.H. Castro Neto, *Biased Bilayer Graphene: Semiconductor with a Gap Tunable by the Electric Field Effect,* Phys. Rev. Lett. **99**, 216802 (2007).

[21]. G. Li, A. Luican and E.Y. Andrei, *Scanning Tunneling Spectroscopy of Graphene on Graphite,* Phys. Rev. Lett. **102**, 176804 (2009).

[22]. G. Trambly de Laissardière, D. Mayou and L. Magaud, *Localization of Dirac Electrons in Rotated Graphene Bilayers,* Nano Lett. **10**, 804 (2010).

[23]. A. Luican, G. Li, A. Reina, J. Kong, R.R. Nair, K.S. Novoselov, A.K. Geim and E.Y. Andrei, *Single-Layer Behavior and Its Breakdown in Twisted Graphene Layers*, Phys. Rev. Lett. **106**, 126802 (2011).

[24]. I. Brihuega, P. Mallet, H. González-Herrero, G. Trambly de Laissardière, M.M. Ugeda, L. Magaud, J.M. Gómez-Rodríguez, F. Ynduráin and J.-Y. Veuillen, *Unraveling the Intrinsic and Robust Nature of van Hove Singularities in Twisted Bilayer Graphene by Scanning Tunneling Microscopy and Theoretical Analysis*, Phys. Rev. Lett. **109**, 196802 (2012).

[25]. L.-J. Yin, J.-B. Qiao, W.-X. Wang, Z.-D. Chu, K.F. Zhang, R.-F. Dou, C.L. Gao, J.-F. Jia, J.-C. Nie and L. He, *Tuning structures and electronic spectra of graphene layers with tilt grain boundaries,* Phys. Rev. B **89**, 205410 (2014).

[26]. A.O. Sboychakov, A.L. Rakhmanov, A.V. Rozhkov and F. Nori, *Electronic spectrum of twisted bilayer graphene,* Phys. Rev. B **92**, 075402 (2015).

[27]. S. Huang, K. Kim, D.K. Efimkin, T. Lovorn, T. Taniguchi, K. Watanabe, A.H. MacDonald, E. Tutuc, and B.J. LeRoy, *Topologically Protected Helical States in Minimally Twisted Bilayer Graphene*, Phys. Rev. Lett. **121**, 037702 (2018).

[28]. F. Zhang, A.H. MacDonald and E.J. Mele, *Valley Chern numbers and boundary modes in gapped bilayer graphene*, Proc. Nat. Acad. Sci. **110**, 10546 (2013).

[29]. F. Zhang, J. Jung, G.A. Fiete, Q. Niu and A.H. MacDonald, *Spontaneous Quantum Hall States in Chirally Stacked Few-Layer Graphene Systems*, Phys. Rev. Lett. **106**, 156801 (2011).

[30]. P. San-Jose and E. Prada, *Helical networks in twisted bilayer graphene under interlayer bias*, Phys. Rev. B. **88**, 121408(R) (2013).

[31]. P. Rickhaus, J. Wallbank, S. Slizovskiy, R. Pisoni, H. Overweg, Y. Lee, M. Eich, M.-H. Liu, K. Watanabe, T. Taniguchi, T. Ihn and K. Ensslin, *Transport Through a Network of Topological Channels in Twisted Bilayer Graphene*, Nano Lett. **18**, 6725 (2018).

[32]. S.G. Xu, A.I. Berdyugin, P. Kumaravadivel, F. Guinea, R. Krishna Kumar, D.A. Bandurin, S.V. Morozov, W. Kuang, B. Tsim, S. Liu, J.H. Edgar, I.V. Grigorieva, V.I. Fal'ko, M. Kim, and A.K. Geim, *Giant oscillations in a triangular network of one-dimensional states in marginally twisted graphene*, Nat. Comm. **10**, 4008 (2019).

[33]. Q. Yao, X. Chen, R. van Bremen, K. Sotthewes, and H.J.W. Zandvliet, *Singularities and topologically protected states in twisted bilayer graphene*, Appl. Phys. Lett. **116**, 011602 (2020).

[34]. J.D. Verbakel, Q. Yao, K. Sotthewes and H.J.W. Zandvliet, *Valley-protected one-dimensional states in small-angle twisted bilayer graphene*, Phys. Rev. B **103**, 165134 (2021).

[35]. Q. Zheng, C.-Y. Hao, X.-F. Zhou, Y.-X. Zhao, J.-Q. He and L. He, *Tunable Sample-Wide Electronic Kagome Lattice in Low-Angle Twisted Bilayer Graphene*, Phys. Rev. Lett. **129**, 076803 (2022).

[36]. J.B. Oostinga, H.B. Heersche, X. Liu, A.F. Morpurgo and L.M.K. Vandersypen, *Gate-induced insulating state in bilayer graphene devices*, Nature Materials **7**, 151 (2008).

[37]. Y. Zhang, T.-T. Tang, C. Girit, Z. Hao, M.C. Martin, A. Zettl, M.F. Crommie, Y. Ron Shen and F. Wang, *Direct observation of a widely tunable bandgap in bilayer graphene*, Nature **459**, 820 (2009).

[38]. W. Yao, S.A. Yang and Q. Niu, *Edge States in Graphene: From Gapped Flat-Band to Gapless Chiral Modes*, Phys. Rev. Lett. **102**, 096801 (2009).

[39]. I. Martin, Y.M. Blanter and A.F. Morpurgo, *Topological Confinement in Bilayer Graphene*, Phys. Rev. Lett. **100**, 036804 (2008).

[40]. L. Ju, Z. Shi, N. Nair, Y. Lv, C. Jin, J. Velasco Jr, C. Ojeda-Aristizabal, H.A. Bechtel, M.C. Martin, A. Zettl, J. Analytis and F. Wang, *Topological valley transport at bilayer graphene domain walls*, Nature **520**, 650 (2015).

[41]. M. Ezawa, *Monolayer topological insulators: silicene, germanene, and stanene*, J. Phys. Soc. Japan **84**, 121003 (2015).

[42]. Y.-H. Shen, J.-D. Zheng, W.-Y. Tong, Z.-Q. Bao, X.-G. Wan, and C.-G. Duan, *Twist-induced quantum valley Hall states in bilayer germanene*, Phys. Rev. B **111**, 075421 (2025).

[43]. E. McCann, D.S.L. Abergel, and V.I. Fal'ko, *Eur. Phys. J. Special Topics* **148**, 91 (2007).

[44]. V.N. Davydov, *Proc. R. Soc. A* **474**, 20180439 (2018).

[45]. N. D. Drummond, V. Z´olyomi, and V. I. Fal'ko, *Electrically tunable band gap in silicene*, Phys. Rev. B **85**, 075423 (2012).

[46]. M. Ezawa, *A topological insulator and helical zero mode in silicene under an inhomogeneous electric field*, New J. Phys., **14**, 033003 (2012).

[47]. H.J.W. Zandvliet, *The quantum valley Hall effect in twisted bilayer silicene and germanene*, J. Phys. Cond. Matt. **37**, 205001 (2025).

[48]. A. Vaezi, Y. Liang, D.H. Ngai, L. Yang, and E.-A. Kim, *Topological Edge States at a Tilt Boundary in Gated Multilayer Graphene*, Phys. Rev. X **3**, 021018 (2013).

[49]. A. Ramires and J.L. Lado, *Electrically Tunable Gauge Fields in Tiny-Angle Twisted Bilayer Graphene,* Phys. Rev. Lett. **121**, 146801 (2018).

[50]. P. Moon and M. Koshino, *Energy spectrum and quantum Hall effect in twisted bilayer graphene*, Phys. Rev. B **85**, 195458 (2012).

[51]. Add ref. releaxation twisted bilayer

[52]. Z. Li, J. Zhuang, L. Chen, Z. Ni, C. Liu, L. Wang, X. Xu, J. Wang, X. Pi, X. Wang, Y. Du, K. Wu, and S.X. Dou, *Observation of van Hove singularities in twisted silicene multilayers,* ACS Cent. Sci. **2**, 517-521 (2016).

[53]. Z. Li, J. Zhuang, L. Wang, H. Feng, Q. Gao, X. Xu, W. Hao, X. Wang, C. Zhang, K. Wu, S.X. Dou, L. Chen, Z. Hu, and Y. Du, *Realization of flat band with possible nontrivial topology in electronic kagome lattice,* Science Adv. **4**, eaau4511 (2018).

[54]. P. Bampoulis, C. Castenmiller, D.J. Klaassen, C. Castenmiller, J. v. Mil, P.L. de Boeij, M. Ezawa and H.J.W. Zandvliet, *Moiré-modulated band gap and van Hove singularities in twisted bilayer germanene*, 2D Materials **11**, 035016 (2024).

[55]. H. Chen, X.-L. Zhang, Y.-Y. Zhang, D. Wang, D.-L. Bao, Y. Que, W. Xiao, S. Du, M. Ouyang, S.T. Pantelides and H.-J. Gao, *Atomically precise, custom-design origami graphene nanostructures*, Science **365,** 1036 (2019).

[56]. J. Kang and O. Vafek, *Analytical solution for the relaxed atomic configuration of twisted bilayer graphene including heterostrain*, Phys. Rev. B **112**, 125138 (2025).

**Appendix A: Bilayer relaxation**

In order to model relaxed bilayer systems, we take the result from Ref. [56] to obtain an analytical displacement field $u^{\pm}(r)$ for the atoms in the top and bottom layers, respectively. This field is given by the expression

$$u^{\pm}(r) = \pm \frac{1}{|G_1|^2} \left\{ \sum_{a=1}^{3} G_a [\zeta_1 \sin(g_a \cdot r) + 2\zeta_3 \sin(2 g_a \cdot r)] + \zeta_2 \sum_{a=1}^{2} (G_a - G_{a+1}) \sin[(g_a - g_{a+1}) \cdot r] \right\}.$$

(A1)

Here, $G_1$ and $G_2$ are the reciprocal lattice vectors of the unrotated layer, $G_3 = -(G_1 + G_2)$, and $G_4 = G_1$. Furthermore, $g_i$ are the moiré reciprocal lattice vectors, which may be obtained from the monolayer lattice vectors through,

$$g_i = R(-\theta/2) G_i - R(+\theta/2) G_i,$$

(A2)

with $R(\theta)$ the rotation matrix in the plane of the bilayer. Furthermore, the (twist-angle-dependent) coefficients $\{\zeta_i\}$ are given by,

$$\zeta_1 = \frac{1}{6\lambda} \left[ -(1 + \alpha\lambda) + \sqrt{(1 + \alpha\lambda)^2 + 12\lambda^2} \right],$$

$$\zeta_2 = \frac{\lambda}{6} \zeta_1 + \frac{\lambda}{3} \frac{c_2}{c_1} - \frac{\lambda}{6} \zeta_1 \left( \zeta_1 + 4 \frac{c_3}{c_1} \right),$$

$$\zeta_3 = \frac{\lambda}{8} \zeta_1 + \frac{\lambda}{4} \frac{c_3}{c_1},$$

(A3)

where,

$$\alpha = \frac{1}{2} + \frac{9}{2} \frac{c_2}{c_1} + 4 \frac{c_3}{c_1}, \qquad \lambda = \frac{c_1}{\mathcal{G}\theta^2}.$$

(A4)

The values of $\{\zeta_i\}$ can then be calculated using the twist angle $\theta$, and the numerical values (for graphene):

$$c_1 = 0.775 \, meV/Å^2, \quad c_2 = -0.071 \, meV/Å^2, \quad c_3 = -0.018 \, meV/Å^2, \quad \mathcal{G} = 9.035 \, eV/Å^2.$$

(A5)

In Fig. (A1), we give an example displacement field for a moiré system with a twist angle $\theta = 1°$. From the displacement field, the new positions of the atoms in the moiré unit cell can be obtained through,

$$r^{\pm}_{relaxed} = r^{\pm}_{unrelaxed} + u^{\pm}(r^{\pm}_{unrelaxed}).$$

(A6)

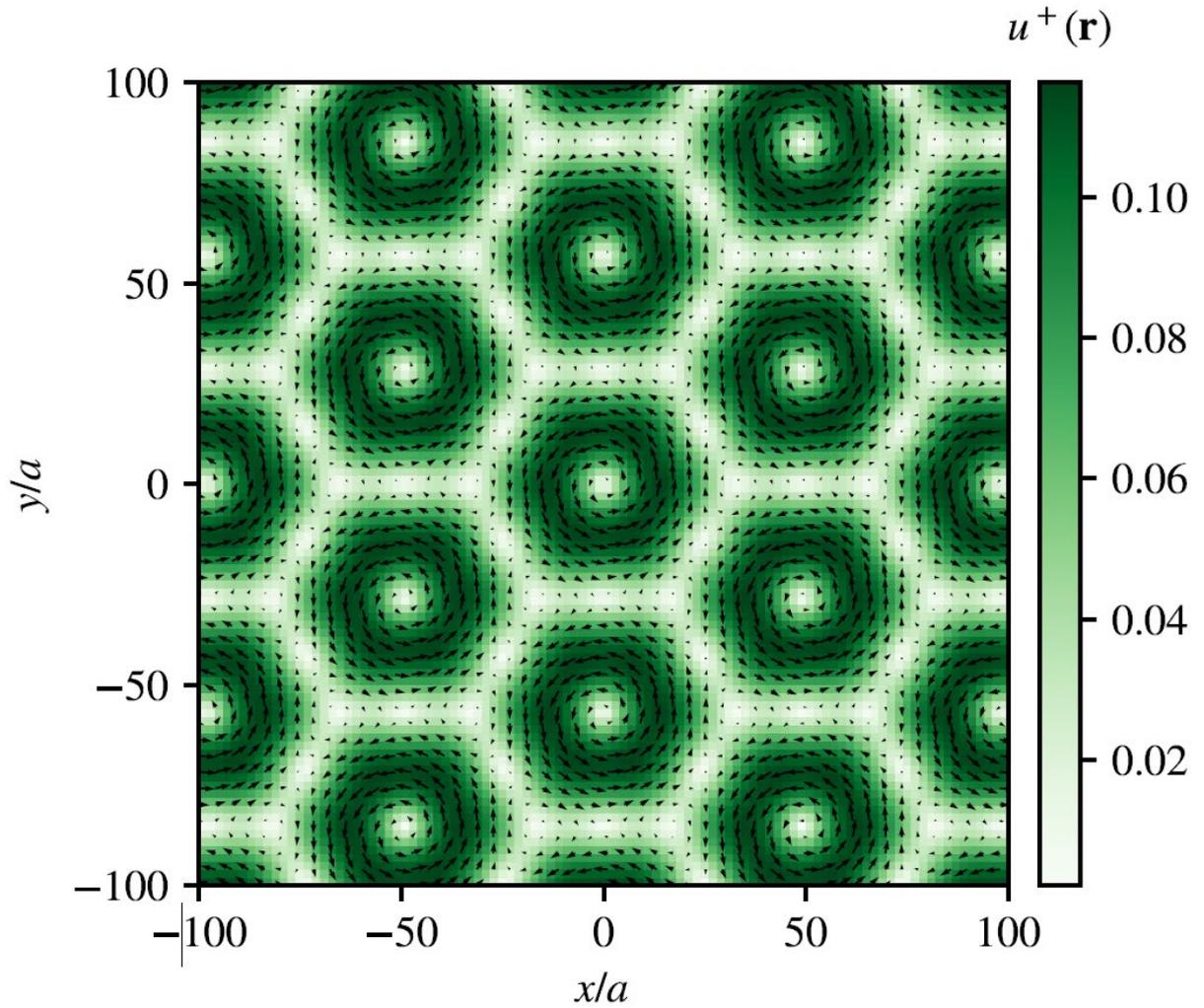

*Figure 6* Relaxation displacement field for a twisted bilayer system with a twist angle.